\pdfoutput=1

\documentclass[preprints,article,accept,moreauthors,pdftex]{Definitions/mdpi}

\firstpage{1} 
\pubvolume{1}
\issuenum{1}
\articlenumber{0}
\pubyear{2021}
\copyrightyear{2020}
\datereceived{} 
\dateaccepted{} 
\datepublished{} 
\hreflink{https://doi.org/} 

\usepackage{amsmath, amssymb}
\usepackage{mathtools}
\usepackage{standalone}
\usepackage{graphicx}
\usepackage[ruled,vlined]{algorithm2e}

\usepackage{tikz}
\usepackage{pgfplots}
\pgfplotsset{width=7cm,compat=1.16}
\usepgfplotslibrary{statistics}

\Title{Private Weakly-Random Sequences from Human Heart Rate for Quantum Amplification}

\TitleCitation{Private weakly-random sequences from human heart rate for quantum amplification}

\Author{
Maciej Stankiewicz$^{1,2}$\orcidA{}, 
Karol Horodecki$^{3,2,}$*\orcidB{}, 
Omer Sakarya$^{3}$\orcidC{}, and 
Danuta Makowiec$^{4}$\orcidD{}
}

\AuthorNames{Maciej Stankiewicz, Karol Horodecki, Omer Sakarya, and Danuta Makowiec}

\AuthorCitation{Stankiewicz, M.; Horodecki, K.; Sakarya, O.; Makowiec, D.}

\address{%
$^{1}$ \quad Institute of Mathematics, Faculty of Mathematics, Physics and Informatics, University of Gdańsk, Wita Stwosza 57, 80-308 Gdańsk, Poland\\
$^{2}$ \quad International Centre for Theory of Quantum Technologies, University of Gdańsk, Wita Stwosza 63, 80-308 Gdańsk, Poland\\
$^{3}$ \quad Institute of Informatics, National Quantum Information Centre, Faculty of	Mathematics, Physics and Informatics, University of Gdańsk, Wita Stwosza 57, 80-308 Gdańsk, Poland\\
$^{4}$ \quad Institute of Theoretical Physics and Astrophysics, Faculty of Mathematics, Physics and Informatics, University of Gdańsk, Wita Stwosza 57, 80-308 Gdańsk, Poland}

\corres{Correspondence: karol.horodecki@ug.edu.pl}

\abstract{
    We investigate whether the heart rate can be treated as a semi-random source with the aim of amplification by quantum devices. We use a semi-random source model called \texorpdfstring{$\varepsilon$}{epsilon}-Santha-Vazirani source, which can be amplified via quantum protocols to obtain fully private random sequence.
	We analyze time intervals between consecutive heartbeats obtained from
	Holter electrocardiogram (ECG) recordings of people of different sex and age.
	We propose several transformations of the original time series into binary sequences.
     We have performed different statistical randomness tests and estimated quality parameters.
	We find that the heart can be treated as good enough, and private by its nature, source of randomness, that every human possesses.
	As such, in principle it can be used as input to quantum device-independent randomness amplification protocols. The properly interpreted \texorpdfstring{$\varepsilon$}{epsilon} parameter can potentially serve as a new characteristic of the human's heart from the perspective of medicine.
}


\begin{document}


\section{Introduction}

Randomness is an essential resource in many applications of everyday life \cite{Bera2017}. As prominent examples, there comes a generation of passwords and tokens for online banking: most transactions are secured by random numbers sent for the user's telephone for authentication purposes. Both public and symmetric key encryption base on a significant amount of randomness. Quantum key distribution developed in the last decades \cite{BB84} is also based on the access to a true random number generator. The choice of measurements applied to devices needs to be unknown to the eavesdropper, i.e., random with respect to her. For these reasons, the privacy of randomness is often the target of hackers' attacks and can be considered one of the ``Achilles' heels'' of security systems. Designing hardware that produces private randomness is a risky task, as recent attacks on such devices show  \cite{Becker2013}.

As a cure for the lack of both reliable and uniform sources of randomness, there comes the theory of the so-called \textit{extractors} \cite{Trevisan, Raz2005}. One can apply a deterministic function for two or more weakly random sources, which extracts (outputs) nearly uniform and private randomness. There are two major models of the weak source
of randomness. The Santha-Vasirani (SV) \cite{SV} source and the $H_{\mbox{min}}$ source \cite{Bell_nonlocality}. It is known that from a \textit{single} such source, amplification is impossible \cite{SV}. As a recent breakthrough, Roger Colbeck and Renato Renner have shown that a single weak source of randomness can be amplified by the use of quantum devices \cite{Colbeck2012, Grudka2014}.
This seminal result has been improved - just two devices are needed for this task both in case of the quantum \cite{ColbeckPHD, Colbeck2011, ChungShiWu, MillerShi} and the so-called no-signaling adversary \cite{Acn2016, Gallego2013, Brando2016, Ramanathan2016}. This effect goes in analogy to the fact that classical randomness extractors amplify the randomness of two \textit{independent} sources of weak randomness.

A significant problem in the use of pairs of devices is the attack, which bases on correlating weak source with device \cite{Wojewodka2017}. One solution that proposes a partial way-out is called ``privatization'' of a weak source of randomness. One chooses as a weak source such emitter of data which is hard to control for anyone, including eavesdropper. Even though the signal comes publicly to the honest parties, due to its \textit{prior} unpredictability to the eavesdropper, its randomness can be quantumly amplified \cite{Max_Rotem}. This solution, however, has a drawback. Between the weak source and quantum amplifier, there is a long chain of detecting devices, each of which can be in principle replaced by a prefabricated source of randomness correlated to the quantum one.

In this manuscript, we propose a novel direction - looking for both hard to predict and hard to replace sources of randomness. Our aim is then amplification of such a source via a quantum device. An example that we focus on in this manuscript is the \textit{heart rate}.

The heart rate has been considered for quite a long as a potential source of randomness in several applications \cite{Seepers2015, Altawy2016, Li2016, Zheng2015}. 
However, it is known that fluctuations of heart rate display scale-invariant non-Gaussian probability density functions \cite{Kiyono.2006} what maps in the observed so-called $1/f$ scaling \cite{Yamamoto.1994} and multifractality \cite{Ivanov.1999} of heart rate. The most pronounce long range correlations are commonly linked to interactions with the respiratory system (3 to 9 beats),  vascular system (9 to 24 beats), and other intrinsic regulatory systems (more than 24 beats) \cite{Shaffer.2017}. Therefore there is discussed whether the heart rate is of the stochastic origin or it is governed by some deterministic nonlinear system \cite{Sassi.2015}.
However, to our knowledge, there is neither proof of its true randomness nor a firm disproof. Both would be interesting practically. The proof would imply a reliable source of weak randomness. Simultaneously, disproof would potentially lead to fast (compressed) heart rate data transfer. This feature would be important in case of an emergency when a patient's data are being sent to the hospital.  Although this type of source is treated as random in several applications, it has been checked recently that raw data from heart rate repositories does not pass tests for randomness \cite{OrtizMartin2018}. However, since we aim in amplifying the randomness of the heart rate, we don't need it to be uniformly random from the beginning.

We perform tests to check to what extent heart rate can be treated as SV-source. One can not directly perform the test of randomness due to the nature of the definition of SV-source. We perform partial tests. After suitable modifications, known as ``cutting out trends'' in medical language, this source, indeed, bares SV-randomness features. We also provide software for testing data for the SV conditions, which may be of independent interest.

Notably, there are classical \textit{extractors} \cite{Raz2005, Chattopadhyay2019} which are applicable in quantum regime. These hash functions output randomness secure from quantum adversary \cite{RPSV2016} when fed with two \textit{independent} sources of weak randomness. Given that it is hard to build a device correlated to someone's heart rate, these extractors are applicable in our context.

The solution we propose is \textit{ecological} as opposed to, e.g., random number generators based on radioactive decay \cite{RANDy}.
It is also cheap and easily accessible due to the rapid development of wearable electronic devices.
One could consider other bio-inspired
sources of randomness, but the heart-based seems to be the easiest to detect \cite{Pawel}. Therefore, we push further the limit of \cite{Aguilar2016}, where it was noted that private randomness is always present in the message. Here we report that the one that is present in the message sender herself can be amplified. What is important, we base on the randomness on which the sender's consciousness has a small impact. Indeed, humans are known to create not random sources when they, e.g., type \cite{Kahn}.

While a low data rate can be seen as a drawback, an option is always randomness expansion \cite{MillerShi}. Amplification results in a small amount of high-quality randomness called a seed. The expansion method uses the seed to generate an arbitrarily long random sequence.

As an interesting side effect, our research can also lead to new findings that are interesting from the medical point of view. Indeed, we present certain cryptographic parameters for healthy volunteers. Comparing this further with the ones obtained from persons with particular diseases can potentially lead to a novel characterization of the latter. 
As a method of pre-processing of the heartbeat, we propose looking for the specific periodic behavior in the data. In this approach, we ``split'' the signal into two sequences, a periodic one and a more random one. The
periodicity can be stronger or weaker. Its strength in the case of patients with certain diseases may potentially better describe particular illnesses.
 
\section{Materials and Methods}

We focus on a model of semi-random sources called Santha-Vazirani source \cite{SV}. An $\varepsilon$-Santha-Vazirani ($\varepsilon$-SV) source is characterized as follows. 
Let \( S = (S_1, S_2, \ldots) \) be a source described by sequence of binary random variables $S_i$. By $s = (s_1, s_2, \ldots)$ we will denote arbitrary long bit string produced by the source $S$ (realization of the source and therefore random variables).
We say that the source is an $\varepsilon$-SV if for all \(i\) 

\begin{equation}
    \label{eq:sv}
	\frac{1}{2}-\varepsilon \leq P(S_{i+1}|S_i, \ldots, S_1, E) \leq \frac{1}{2} + \varepsilon
\end{equation}

where \(E\) represents an arbitrary random variable prior to \(S_1\) that can influence the source. It is easy to see that for \( \varepsilon = 0 \) the bits are fully random and for \( \varepsilon = 1/2 \) they can even be deterministic. 

The most straightforward example of the SV source is the source of tossing of the fake coin with distribution $\{1/2 +\varepsilon, 1/2 - \varepsilon\}$. Moreover, the $\varepsilon$-SV source has been characterized in \cite{Grudka2014} as a mixture of specific permutations of this distribution. As such, it does not pass standard randomness tests, such as Dieharder \cite{Dieharder}. However, as we argue, it appears to be random enough to be amplified to uniformly random bit-string if combined with another independent device. E.g., one can use the amplification given in \cite{Max_Rotem} (although we don't need privatization of an SV source, as the SV source of our choice is practically private).

\subsection{Used data and their preprocessing}\label{sec:preprocessing}

Twenty-four-hour Holter ECG recordings during a normal sleep-wake cycle were obtained from healthy volunteers without any known cardiac history. The Holter recordings were analyzed using Del Mar Reynolds Impresario software and screened for premature, supraventricular and ventricular beats, missed beats, and pauses. These data were annotated correspondingly by the automatic system. The entries which were not annotated as normal (N) were neglected.

While such pre-processing have little physiological significance, it is satisfactory for the cryptographic purpose. We will refer to these data as pre-processed for cryptographic purposes.

The data pre-processed for cryptographic purposes were organized as follows. Each person has a separate text file with sex, age, and start time of measurement encoded into the file name. In each file, after the header, there are four columns: number of the observation, time of observation, length of RR interval, and annotation. Since Del Mar software used to obtain data has a 128Hz sampling frequency (which means approximately 8ms resolution) the entries in the RR interval column are represented as a limited set of rational numbers.

In order to obtain a physiological meaning to the data, the signals had to be pre-processed by the experienced cardiologist. It occurs that the activity of the heart during the day is affected by some external factors due to brain stimulation via interactions with the environment which makes the signal analysis extremely difficult. For this reason, only the nocturnal part of the Holter record has been thoroughly corrected manually by cardiologists and annotated correspondingly. 

The hours of sleep were identified for each signal individually in order to detect properly the day-night transition. A six-hour period, covering the longest RR-intervals, was extracted as the nocturnal period.  Perturbations in signals - artifacts or not normal-to-normal RR-intervals were edited as follows. Perturbations consisting of less than five consecutive RR-intervals were replaced by the median estimated from the last seven normal RR-intervals. Other perturbations were deleted. Eventually, the nocturnal signals were constructed from at least 20000 RR intervals. 

We will denote pre-processed sequence of RR intervals as $\{d_i\}_{i=1}^n$. We will refer to this data as to \textit{manually pre-processed}

\subsubsection{Discretization}\label{sec:discretization}

Since, for purpose or randomness amplification (and testing SV source parameter), we need binary sequence we have to apply some form of discretization. We have tested three classes of parameterized function that maps a string of rational numbers $\{d_i\}_{i=1}^n$ to a binary sequence $\{s_i\}_{i=1}^n$.

The first one assigns deceleration of the heart rate to 0 and acceleration to 1 in the following way

\begin{equation}
	s_i = 
	\begin{cases}
		0 :& d_i \geq d_{i-1} + \eta_1\\
		1 :& \text{otherwise}
	\end{cases},
\end{equation}

where $\eta_1$ is an offset parameter.
The second one outputs 0 if the change in heart rate is rapid (above threshold $\eta_2$) and 1 if it is slow using the equation 

\begin{equation}
s_i = 
\begin{cases}
0 :& |d_i - d_{i-1}| \geq \eta_2\\
1 :& \text{otherwise}
\end{cases}.
\end{equation}

The last one, in some sense, take into consideration the monotonicity of three consecutive heart bits in the following way 

\begin{equation}
s_i = 
\begin{cases}
0 :& d_i \geq d_{i-1} \geq d_{i-2} \lor d_i \leq d_{i-1} \leq d_{i-2} \\
1 :& \text{otherwise}
\end{cases}.
\end{equation}

During the initial phase of our experiment, we discovered that the best results are obtained using the first method of discretization with $\eta_1 = 0$. Therefore, the results presented in the remaining part of the paper will use this method.

\subsubsection{Cutting out trends}\label{sec:cutting}

Let us note first, that the data from the heart have partially periodic behavior by the nature of the source: after several consecutive accelerations, there needs to come, sooner or later, several slowdowns. This implies, that this source can not satisfy the SV condition: a too long sequence of $1$s denoting accelerations can not appear in the data. On the other hand in $n$ long bit-string $\varepsilon$-SV source it has probability $(\frac{1}{2}-\varepsilon)^{n}>0$. However, we can modify the source by cutting out the acceleration and deceleration parts. This is known as \textit{cutting out trends}. This approach is parametrized by pair of natural numbers $(i,j)$. It means that
we cut out $i$ consecutive signals
of acceleration and the first next $j$ consecutive signals of slow down, after
which we look for the next $i$ accelerations and $j$ decelerations and so on through the whole sequence. In our experiment, we consider only the case $i=j \in \{3,...,6\}$.
 
As we will see, this method resulting in lowering the rate of the source yields a more random one.
By doing so we base on the fluctuations of the heart rate. Note that this approach is different from other considered in literature \cite{Bouda2009} that base on fluctuations of the measurement devices such as a camera.
Such methods based on fluctuations or noise of the signal recorded by the measurement devices can be prone to attacks that use malicious measurement devices. On the other hand, our approach uses the main part of the signal (heart rate) and presumably can be resistant to such vectors of attack. It can be done by using several different measurement devices from different manufacturers and comparing the results.

\subsection{Randomness testing method details}\label{sec:randtestmethod}

In this section, we will describe the idea behind our testing method called SVTest implemented in the software we use.

Our goal is to estimate $\varepsilon$ from Eq.\ (\ref{eq:sv}). We will do it by estimating values $\varepsilon_h$ where $h \in \{0, \ldots, h_{\mathrm{max}}\}$ are history lengths (condition lengths). We will postpone reasoning how to choose $h_{\mathrm{max}}$ to the later part of the paper. Using the definition of conditional probability and more verbose notation we can rewrite Eq.\ (\ref{eq:sv}) to obtain value $\varepsilon_h$ in explicit way

\begin{equation}
	\label{eq:SV2}
	\varepsilon_h \coloneqq \max_{s_{n-h+1}, \ldots, s_{n+1} \in \{ 0,1 \}}
	\quad\left \lvert \frac{P(s_{n+1}, s_n, s_{n-1}, \ldots, s_{n-h+1})}{P(s_n, s_{n-1}, \ldots, s_{n-h+1})} - \frac{1}{2} \right \rvert.
\end{equation}

We should point out here that the above definition makes sense only for random number generating devices (that are modeled by probability distributions). Since we treat the device as a black box we have only access to some binary sequence $s = (s_1, s_2, \ldots, s_n)$ that is outputted by the device (that can be seen as a realization of the device's probability distribution). Because of this, our approach to approximate $\varepsilon_h$ is to use frequencies to estimate probabilities from Eq.\ (\ref{eq:SV2}). By doing this we obtain

\begin{equation}
	\label{eq:epsilonEstimation}
	\tilde{\varepsilon}_h \coloneqq 
	\max_{w_h} \left \lvert \frac{\frac{|s|_{w_h}}{n-h}}{\frac{|s|_{w_h'}}{n-h+1}} - \frac{1}{2} \right \rvert \mathrel{\mathop{\approx}\limits_{n\to \infty}}
	\max_{w_h} \left \lvert \frac{|s|_{w_h}}{|s|_{w_h'}} - \frac{1}{2} \right \rvert
\end{equation}

where $w_h \coloneqq (x_1, x_2, \ldots, x_h)$ is binary sequence, $w' \coloneqq (x_1, x_2, \ldots, x_{h-1})$, and $|a|_b$ denotes number of occurrence of substring $b$ in string $a$.
Here, by $\tilde{\epsilon}_h(s_n)$ we denote
the experiment-driven value to make
it distinct from the theoretical one$\epsilon_h$.

\subsubsection{Similar tests used in previous research} 
	
The so-called serial test from the seminal NIST test suite \cite{NIST} uses a similar approach to our method. The serial test is focused on checking the frequency of all overlapping $h$-bit sequences. The three main differences are that:
\begin{itemize}
	\item frequencies are calculated instead of conditional values, 
	\item average square deviation is used instead of maximal absolute deviation,
	\item cyclic approach is used at the end of the sequence.
\end{itemize}
Also, the NIST test suite was developed to check pseudo-random sequences to use them directly in classical algorithms. That approach demands the sequence to be almost perfectly random. On the other hand, since we apply the quantum randomness amplification method, it is enough that the sequence is partially random assuming that we know the threshold $\varepsilon$.

Another, quite a similar testing method was presented by Martinez et.\ al.\ \cite{Martnez2018} that is based on the Borel-normality criterion \cite{Borel-normality} but in their approach overlapping is not used. For us, it is interesting, that they claim that the longest history length used should be of the order of $\log_2(\log_2(n))$.

\subsection{Finite effects}

\begin{Remark}
For the sequence of data $s_n$ of length $n$, the maximal length of history  $h$ that satisfies $\varepsilon$-SV condition (\ref{eq:sv}) with $\varepsilon < 1/2$ satisfies:

\begin{equation}
    h \leq \log (n) -1
    \label{eq:trivial_bound}
\end{equation}

\label{obs:trivial}
\end{Remark}

\begin{proof}
We note that the condition (\ref{eq:sv}) implies
that $2^{h+1} \leq n-(h+1)$. This is because it implies that all sequences of length $h+1$ (of history and one bit history of which is considered) should appear in the sequence $x_n$. The term $h+1$ on r.h.s. appears because the last position at which such sequence can start
is $n-h-1$, as it has length $h+1$ and should be a subsequence of $x_n$.
Taking logarithm on both sides we obtain
$h+1 \leq \log(n-h-1)\leq \log n$, hence the assertion follows
\end{proof}

\subsection{Identifying \texorpdfstring{$\varepsilon$}{epsilon} from a realization of a source of weak randomness}\label{sec:singleEpsilon}

We propose a method to estimate $\varepsilon$ of the underlying SV source given \textit{finite} realization of this source. It is independent of the type of the source, hence may be of interest itself, however, the argument is heuristic.

First, we note, that given a finite sequence of data we can not expect $\varepsilon$-SV conditions to be satisfied for all lengths of histories. A trivial observation is that the maximal history length $h$ in a sequence of length $n$ satisfies

\begin{equation}
    h \leq \log(n) -1
\end{equation}

(see Observation \ref{obs:trivial}).

One can ask if there exists a realization which saturates the above inequality. We give the affirmative answer by observing that the so-called De Bruijn sequences have this property \cite{DeBruijn}. For any $n=2^k$, the De Bruijn sequence $d_k$ of length $n$ satisfies ideally, i.e., with $\varepsilon=0$ the $\varepsilon$-SV conditions given in Eq.\ (\ref{eq:sv}) up to history of length $\log n -1=k-1$. 
Unlike in our considerations, in De Bruijn's construction one assumes \textit{cyclic} boundary conditions, i.e., that $d_k[n]$ is neighbour of $d_k[1]$.
We give the first two exemplary sequences below.

\begin{align}
    d_2&=0011 \\
    d_3&=00010111 
\end{align}

For example, in the sequence $d_2$ ``0'' appears the same number of times as ``1''. The same holds for sequences of pairs of bits, each appearing once (sequence ``10'' is obtained from the cyclic condition).

For some sequences, one can however observe non-zero $\varepsilon$ for histories of less length than $\log (n)-1$. However one needs to attribute \textit{single} $\varepsilon$ to the source. It is rather plausible, that the longest histories correspond to events with no statistical meaning as they appear only a few times (or even once) in total sequence. We, therefore, propose to use the \textit{weighted} average, defining $\varepsilon$ for the sequence $s_n$ as

\begin{equation}
\label{eq:epsilon}
    \tilde\varepsilon(s_n):= \frac{1}{w(\lfloor\log_2(n)\rfloor-1)}\sum_{i=0}^{\lfloor\log_2(n)\rfloor-1} \frac{\tilde\varepsilon_i(s_n)}{(i+1)}
\end{equation}

with $w(h)=\sum_{i=0}^{h} \frac{1}{i+1}$.

\subsection{Software description}

\begin{algorithm}
\SetAlgoLined
\KwData{Annotated data file from Holter device, maximal history length parameter $h_{max}$}
\KwResult{Sequence of $\varepsilon_i$ for $i \in \{0, \ldots, h_{max}\}$}
\nl Read appropriate RR intervals from data file\;
\nl Generate binary sequence from RR intervals using chosen discretization\;
\nl Optionally: perform cutting out trends subroutine\;
\nl Count the number of occurrences of each binary substring of length up to $h_{max} + 1$\;
\nl Calculate estimated epsilons given in Eq.\ (\ref{eq:epsilonEstimation})\;
\caption{Estimation of epsilons}
\end{algorithm}

In the first step, we read into memory sequence of RR intervals $\{d_i\}_{i=1}^n$. Furthermore, we only take into account the ones that are valid according to annotations (see the beginning of Section \ref{sec:preprocessing} for details).

In the second step, the sequence of rational numbers $\{d_i\}_{i=1}^n$ is changed into binary string $\{s_i\}_{i=1}^n$ according to chosen discretization. Exemplary classes of functions that can be used to diesesscretize are discussed in Section \ref{sec:discretization}.

In the third step, the so-called cutting out trends is performed. It is not mandatory since it reduces the number of bits, but in some cases, it can be beneficial due to smaller epsilons in the shorter resulting sequence. We discuss this method in Section \ref{sec:cutting}.

In the fourth step, we count the number of occurrences of each binary string of length not larger than $h_{max} + 1$ as a consecutive substring of our binary data string. It is done in such a way that substrings can overlap. We do not assume cyclicity, i.e., the end of the data string is not considered as following the beginning of it. For example in the sequence 001011011 substring ``0'' appears four times, substring ``1'' five times, substring ``00'' one time, substring ``01'' three times, substring ``10'' two times, substring ``11'' two times, substring ``000'' zero times, and so on.

In the fifth step, we use appropriate numbers of occurrences from the previous step to calculate epsilons according to the formulas from Section \ref{sec:randtestmethod}. At this point the program outputs sequences $\{\varepsilon_i\}_{i=1}^{h_{max}}$.

An additional step is to obtain single epsilon from the above sequence of epsilons. Since, it was easier to analyze epsilons' behavior we perform that step separately, outside of our program. Our suggested way of calculating single, final epsilon is described in Section \ref{sec:singleEpsilon}.

\subsubsection{Implementation}

Our software is implemented in the C programming language. We used several optimization features to speed up the analysis. Since we are interested in overlapping substrings we treated sub-sequences as natural numbers and used bit shift operation to process the next bit. We also count the number of occurrences of each substring only once and use it to calculate $\varepsilon_i$ for all $i$ at the very end instead of processing each history length separately. Furthermore, since a number of occurrences of the sequences starting with ``1'' can be deduced from appropriate sequences that start with ``0'' and previously calculated shorter ones we can omit it reducing running time and amount of used memory.

Such optimization was sufficient because of (the characteristic for our research) type and a moderate amount of data.  On the other hand, there are some other features that can be implemented in future versions of the software if needed. For example, for extremely large data sets reading it whole at once can be problematic because of the limited amount of memory. In such a case, reading files in smaller portions can be a better solution. Also, a multi-threaded version of the program could be considered. Finally, we can even think of reading data from the stream and estimating epsilons on the fly for example in real-time applications.

\section{Results}

We will present experiments taking two points of view: cryptographic and medical. For medical analysis, we will consider data that are pre-processed by medical experts (see Section \ref{sec:preprocessing}). For cryptographic purposes, we consider raw data, with less physiological meaning. The latter approach is justified, as, in a potential application, there will be no place for an expert to pre-process the raw data before their randomness is quantumly amplified. Indeed, it would be not only impractical but also against the approach of device-independent processing which assumes no trust to third parties.

\subsection{Identifying \texorpdfstring{$\varepsilon$}{epsilon} for exemplary raw data for cryptographic purpose}\label{sec:crypto}

The heart rate data, which we consider here come from $8$-hour long recording of many persons (64 women and 51 men). We, therefore, lack a large bulk of data. To overcome this, we consider a merged file of data from 118 persons as if they come from a much longer period for a single person. In Table \ref{tab:epsilons} we show the values of the $\varepsilon$ based on the merged file.

\begin{specialtable}[H] 
\centering
\captionsetup{justification=centering}
    \caption{Values of epsilons.\label{tab:epsilons}}
    \begin{tabular}{|c|c|c|c|c|}
        \hline
        size (bits) & 1 191 328 & 488 968 & 241 072 & 120 997 \\\hline
        $\varepsilon$ & 0.20972 & 0.190349 & 0.201891 &0.185005\\\hline
        pattern of cut & (3,3) & (4,4) & (5,5) & (6,6) \\\hline
        $\varepsilon_{cut}$ & 0.134322 & 0.105582 & 0.104898 & 0.0934031 \\\hline
        $\varepsilon -\varepsilon_{cut}$ & 0.0753987 & 0.0847669 & 0.0969931 & 0.0916016 \\\hline
    \end{tabular}
\end{specialtable}


In the table we compare an $\varepsilon$ from the raw data $x_n$ of size $n$, with data $c_m$ obtain from some $m>n$ size after cutting out trends. Here $m$ is large enough so that after cutting out, the remaining data has the same size $n$ as raw data.  
We further show also the values of $\varepsilon$ for different persons. We focus on cutting out trends with pattern $(3, 3)$. Although cutting out larger patterns $(i, i)$ yields better $\varepsilon$, we focus on this case because the output string
is too short for $i > 3$.

\begin{figure}[htbp]
    \centering
    \begin{tikzpicture}
    \begin{axis}[boxplot/draw direction=y,x axis line style={opacity=0},axis x line*=bottom,axis y line=left,enlarge y limits,ymajorgrids,xtick={1,2,3,4,5,6,7},xticklabels={20, 30, 40, 50, 60, 70, 80},]

    \addplot+ [
        boxplot prepared={
            lower whisker=  0.122324589075867,
            lower quartile= 0.171124694646745,
            median=         0.181034274929185,
            upper quartile= 0.191959089604186,
            upper whisker=  0.249482657668713,
        },
    ] 
    coordinates{};
    
    \addplot+ [
        boxplot prepared={
            lower whisker=  0.162706158831961,
            lower quartile= 0.176182337877612,
            median=         0.195214887292934,
            upper quartile= 0.204545305200677,
            upper whisker=  0.230161514450493,
        },
    ] 
    coordinates{};
    
    \addplot+ [
        boxplot prepared={
            lower whisker=  0.171622925338337,
            lower quartile= 0.20187780498547,
            median=         0.210714147415096,
            upper quartile= 0.236317163157004,
            upper whisker=  0.253372491252736,
        },
    ] 
    coordinates{};
    
    \addplot+ [
        boxplot prepared={
            lower whisker=  0.184925586737257,
            lower quartile= 0.214259588749968,
            median=         0.24359359076268,
            upper quartile= 0.267289199679807,
            upper whisker=  0.290984808596934,
        },
    ] 
    coordinates{};
    
    \addplot+ [
        boxplot prepared={
            lower whisker=  0.180203474240623,
            lower quartile= 0.212809395288928,
            median=         0.230768189499045,
            upper quartile= 0.254806747001238,
            upper whisker=  0.29707949407007,
        },
    ] 
    coordinates{};
    
    \addplot+ [
        boxplot prepared={
            lower whisker=  0.213504173970161,
            lower quartile= 0.234410950153037,
            median=         0.253244543370774,
            upper quartile= 0.271294738833009,
            upper whisker=  0.290712012676134,
        },
    ] 
    coordinates{};
    
    \addplot+ [
        boxplot prepared={
            lower whisker=  0.25552361143159,
            lower quartile= 0.269288269571126,
            median=         0.272903058017117,
            upper quartile= 0.280391478242905,
            upper whisker=  0.289200062812904,
        },
    ] 
    coordinates{};
        
    \end{axis}
    \end{tikzpicture}
    
    \caption{Values of $\varepsilon$ for the group of men, aged $21$--$88$, shown in increasing order of age. The bottom line of the rectangular shape denotes the first quartile, the top one denotes the third quartile, and the middle one denotes the second quartile (median). The minimal and maximal values (the $0$th and $4$th quartiles) are depicted as top and bottom whiskers for each group of age.}
    \label{fig:menFull}
\end{figure}
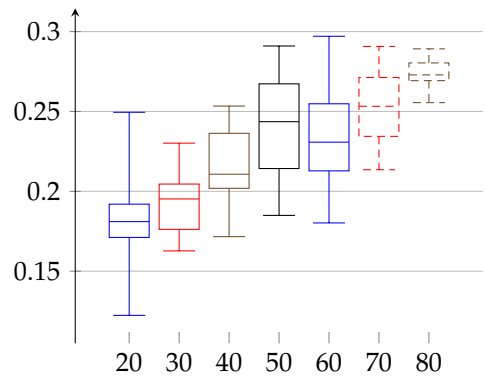

We are ready to describe the experimental results of epsilon for two groups of volunteers. 
The first group consists of $66$ of women aged $19$--$89$ years.
The second consists of $52$ men aged $21$--$88$ years. We divided each group into subgroups of persons with age belonging to
interval $[10*j,10*(j+1))$ for $j\in\{1,...,8\}$ for women and $[10*k,10*(k+1))$ with $k \in \{2,...,8\}$. 
The age groups consisted of 10: 3, 20: 9, 30: 11, 40: 7, 50: 6, 60: 9, 70: 9, and 80: 12 elements in case of women and 20: 11, 30: 10, 40: 8, 50: 3, 60: 5, 70: 9, and 80: 6 in case of men.
On Figures \ref{fig:womenFull} and \ref{fig:menFull} we describe the computed $\epsilon$ 
according to definition given in Eq.\ (\ref{eq:epsilon}). In case of 
women the value of $\epsilon^{(w)} \in [0.17588, 0.35052]$ and the median has growing tendency. The data of women with age below $20$ may have low statistical meaning due to low number of persons in this group. These trends should be confirmed with more persons in a given range of the age.
Similarly, in case of men the median of epsilon also has increasing tendency with age and is in the range $\epsilon^{(m)} \in [0.12232, 0.29707]$ (note that the values for men with age in the rage $50$-$59$ have low statistical meaning, due to low number of volunteers in this age). 
In the second experiment, depicted on
Fig. \ref{fig:womenNormalized} and Fig. \ref{fig:menNormalized}, we show the results 
for the $\epsilon$ achieved for the same initial amount of data. In this case $\epsilon^{(w)} \in [0.17262, 0.35135]$ and $\epsilon^{(m)}\in [0.13295, 0.29990]$. In that case, the value of median of $\epsilon$ has growing tendency for both women and men. 

In the final experiment (see Fig. \ref{fig:women_after_cut2} and \ref{fig:men_after_cut2} respectively) we apply the post-processing of cutting trends
described in the above. In result $\epsilon$ lowers in both cases to reach the ranges $\epsilon^{(w)} \in [0.08539, 0.23153]$ for women and $\epsilon^{(m)} \in [0.08327, 0.20838]$ for men.

\begin{figure}[htbp]
    \centering
    \begin{tikzpicture}
    \begin{axis}[boxplot/draw direction=y,x axis line style={opacity=0},axis x line*=bottom,axis y line=left,enlarge y limits,ymajorgrids,xtick={1,2,3,4,5,6,7,8},xticklabels={10,20, 30, 40, 50, 60, 70, 80},]
    \addplot+ [
        boxplot prepared={
            lower whisker=0.185477371404697,
            lower quartile=0.200005713140641,
            median=0.214534054876586,
            upper quartile=0.224240071265559,
            upper whisker=0.233946087654533,
        },
        ] 
        coordinates{};
      
    \addplot+ [
        boxplot prepared={
            lower whisker=0.175882339963827,
            lower quartile=0.185477371404697,
            median=0.187275002909978,
            upper quartile=0.255292207247372,
            upper whisker=0.257176287131844,
        },
    ] 
    coordinates{};

    \addplot+ [
        boxplot prepared={
            lower whisker=0.192036123838595,
            lower quartile=0.210415513894215,
            median=0.224502536267692,
            upper quartile=0.239833834633318,
            upper whisker=0.265726762974106,
        },
    ] 
    coordinates{};

    \addplot+ [
        boxplot prepared={
            lower whisker=0.188542669128918,
            lower quartile=0.209945187924846,
            median=0.232658214072648,
            upper quartile=0.244407469938186,
            upper whisker=0.25614925719815,
        },
    ] 
    coordinates{};
    
    \addplot+ [
        boxplot prepared={
            lower whisker=0.187228722574604,
            lower quartile=0.216894553911961,
            median=0.226994634514138,
            upper quartile=0.24245961764566,
            upper whisker=0.27260417226022,
        },
    ] 
    coordinates{};
    
    \addplot+ [
        boxplot prepared={
            lower whisker=0.206066715924408,
            lower quartile=0.229609803764056,
            median=0.248484080715722,
            upper quartile=0.266823000344066,
            upper whisker=0.318620586399243,
        },
    ] 
    coordinates{};
    
    \addplot+ [
        boxplot prepared={
            lower whisker=0.225361674729078,
            lower quartile=0.231531313113033,
            median=0.279539182001058,
            upper quartile=0.29117794239339,
            upper whisker=0.350524603419143,
        },
    ] 
    coordinates{};
    
    \addplot+ [
        boxplot prepared={
            lower whisker=0.238848339577059,
            lower quartile=0.261977122457775,
            median=0.269731953477909,
            upper quartile=0.288090727561208,
            upper whisker=0.331128124319718,
        },
    ] 
    coordinates{};
        
    \end{axis}
    \end{tikzpicture}
    
    
    \caption{Values of $\varepsilon$ for the group of women, aged $19$--$89$, shown in increasing order of age, (without cutting trends). The bottom line of the rectangular shape denotes the first quartile,  the top one denotes the third quartile, and the middle one denotes the second quartile (median). The minimal and maximal values (the $0$th and $4$th quartiles) are depicted as top and bottom whiskers for each group of age.
    }
    \label{fig:womenFull}
\end{figure}
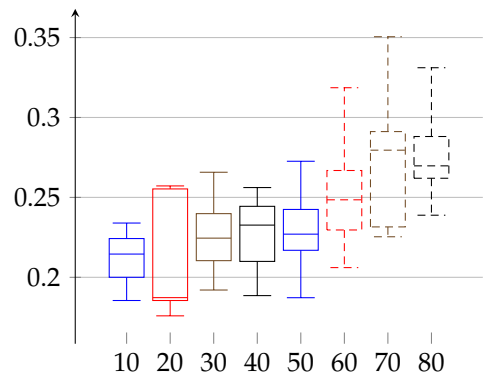

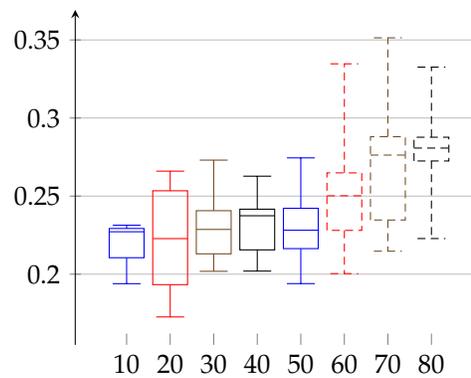
\begin{figure}[htbp]
    \centering
    \begin{tikzpicture}
    \begin{axis}[boxplot/draw direction=y,x axis line style={opacity=0},axis x line*=bottom,axis y line=left,enlarge y limits,ymajorgrids,xtick={1,2,3,4,5,6,7,8},xticklabels={10,20, 30, 40, 50, 60, 70, 80},]

    \addplot+ [
        boxplot prepared={
            lower whisker=  0.193905819434031,
            lower quartile= 0.210534186028467,
            median=         0.227162552622904,
            upper quartile= 0.229252904545089,
            upper whisker=  0.231343256467275,
        },
    ] 
    coordinates{};
    
    \addplot+ [
        boxplot prepared={
            lower whisker=  0.172624122329441,
            lower quartile= 0.193291589298413,
            median=         0.222776545056381,
            upper quartile= 0.253406247888923,
            upper whisker=  0.265978945246232,
        },
    ] 
    coordinates{};

    \addplot+ [
        boxplot prepared={
            lower whisker=  0.201839435028995,
            lower quartile= 0.213027529075866,
            median=         0.22865692965454,
            upper quartile= 0.240649956984291,
            upper whisker=  0.273044018856613,
        },
    ] 
    coordinates{};

    \addplot+ [
        boxplot prepared={
            lower whisker=  0.201970251724534,
            lower quartile= 0.215561806616519,
            median=         0.237412917752964,
            upper quartile= 0.241621664115096,
            upper whisker=  0.262755504123101,
        },
    ] 
    coordinates{};
    
    \addplot+ [
        boxplot prepared={
            lower whisker=  0.193947490557113,
            lower quartile= 0.216362557971779,
            median=         0.228142491065667,
            upper quartile= 0.242192137810525,
            upper whisker=  0.27447722035101,
        },
    ] 
    coordinates{};
    
    \addplot+ [
        boxplot prepared={
            lower whisker=  0.20036627219107,
            lower quartile= 0.228043455684057,
            median=         0.250192731423441,
            upper quartile= 0.264859905510739,
            upper whisker=  0.334698338206546,
        },
    ] 
    coordinates{};
    
    \addplot+ [
        boxplot prepared={
            lower whisker=  0.214728169453414,
            lower quartile= 0.234689512134326,
            median=         0.276266729056383,
            upper quartile= 0.288019750342048,
            upper whisker=  0.351356813229199,
        },
    ] 
    coordinates{};
    
    \addplot+ [
        boxplot prepared={
            lower whisker=  0.222813815335766,
            lower quartile= 0.272486792453226,
            median=         0.280758571086935,
            upper quartile= 0.287696606679046,
            upper whisker=  0.332583554386759,
        },
    ] 
    coordinates{};
        
    \end{axis}
    \end{tikzpicture}
    
    
    \caption{Values of $\varepsilon$ for the group of woman aged $21$--$88$ shown in increasing order of age for the same length of data (cut to the minimal length among all persons). Depiction of statistical quantities ($0$--$4$th quartile) as described in caption of Fig \ref{fig:womenFull}. }
    \label{fig:womenNormalized}
\end{figure}

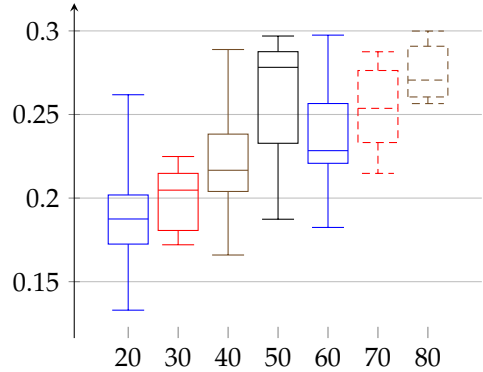
\begin{figure}[htbp]
    \centering
    \begin{tikzpicture}
    \begin{axis}[boxplot/draw direction=y,x axis line style={opacity=0},axis x line*=bottom,axis y line=left,enlarge y limits,ymajorgrids,xtick={1,2,3,4,5,6,7},xticklabels={20, 30, 40, 50, 60, 70, 80},]

    \addplot+ [
        boxplot prepared={
            lower whisker=  0.132951832681267,
            lower quartile= 0.172483202341895,
            median=         0.1875009662206,
            upper quartile= 0.201863179179418,
            upper whisker=  0.261817206148533,
        },
    ] 
    coordinates{};

    \addplot+ [
        boxplot prepared={
            lower whisker=  0.172067677310934,
            lower quartile= 0.180628093689826,
            median=         0.204747816598571,
            upper quartile= 0.214780143438723,
            upper whisker=  0.224862708413701,
        },
    ] 
    coordinates{};

    \addplot+ [
        boxplot prepared={
            lower whisker=  0.165948171051773,
            lower quartile= 0.203969187926598,
            median=         0.216638808445093,
            upper quartile= 0.238264253825593,
            upper whisker=  0.288868942536743,
        },
    ] 
    coordinates{};
    
    \addplot+ [
        boxplot prepared={
            lower whisker=  0.187350766735431,
            lower quartile= 0.232781485360193,
            median=         0.278212203984955,
            upper quartile= 0.287574454355583,
            upper whisker=  0.296936704726212,
        },
    ] 
    coordinates{};
    
    \addplot+ [
        boxplot prepared={
            lower whisker=  0.18243930822309,
            lower quartile= 0.220801170631115,
            median=         0.22836795320864,
            upper quartile= 0.256559392581331,
            upper whisker=  0.297475478925171,
        },
    ] 
    coordinates{};
    
    \addplot+ [
        boxplot prepared={
            lower whisker=  0.214768911208807,
            lower quartile= 0.233224126911023,
            median=         0.253674656422225,
            upper quartile= 0.276311552262877,
            upper whisker=  0.287527137341622,
        },
    ] 
    coordinates{};
    
    \addplot+ [
        boxplot prepared={
            lower whisker=  0.256521080674261,
            lower quartile= 0.260495424411243,
            median=         0.27058366818018,
            upper quartile= 0.29082353896114,
            upper whisker=  0.299909876917606,
        },
    ] 
    coordinates{};
        
    \end{axis}
    \end{tikzpicture}
    
    
    \caption{Values of $\varepsilon$ for the group of man aged $21$--$88$ shown in increasing order of age for the same length of data (cut to the minimal length among all persons). Depiction of statistical quantities ($0$--$4$th quartile) as described in caption of Fig \ref{fig:womenFull}. }
    \label{fig:menNormalized}
\end{figure}

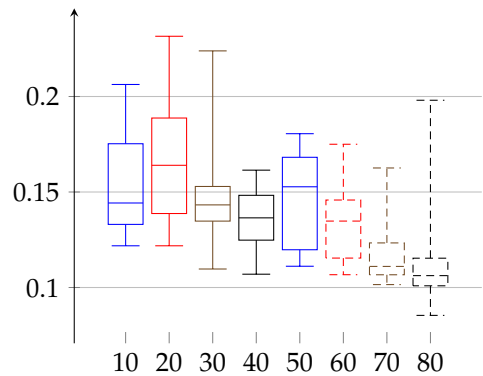
\begin{figure}[htbp]
    \centering
    \begin{tikzpicture}
    \begin{axis}[boxplot/draw direction=y,x axis line style={opacity=0},axis x line*=bottom,axis y line=left,enlarge y limits,ymajorgrids,xtick={1,2,3,4,5,6,7,8},xticklabels={10,20, 30, 40, 50, 60, 70, 80},]
    \addplot+ [
        boxplot prepared={
            lower whisker=0.121839666581445,
            lower quartile=0.133043078829884,
            median=0.144246491078323,
            upper quartile=0.175272011107274,
            upper whisker=0.206297531136225,
        },
        ] 
        coordinates{};
        
    \addplot+ [
        boxplot prepared={
            lower whisker=0.121839666581445,
            lower quartile=0.138732613555103,
            median=0.16398344268052,
            upper quartile=0.18873337591244,
            upper whisker=0.231539049293252,
        },
    ] 
    coordinates{};
    
    \addplot+ [
        boxplot prepared={
            lower whisker=0.109683048632432,
            lower quartile=0.134761596263733,
            median=0.143253535464263,
            upper quartile=0.152896692161722,
            upper whisker=0.223883882792217,
        },
    ] 
    coordinates{};
        
    \addplot+ [
        boxplot prepared={
            lower whisker=0.106964002514966,
            lower quartile=0.124794322171269,
            median=0.136464260516174,
            upper quartile=0.148262979770289,
            upper whisker=0.161437441750362,
        },
    ] 
    coordinates{};
    
    \addplot+ [
        boxplot prepared={
            lower whisker=0.111127644824228,
            lower quartile=0.119767377796055,
            median=0.152735740649418,
            upper quartile=0.168171776870323,
            upper whisker=0.180529706888824,
        },
    ] 
    coordinates{};
    
    \addplot+ [
        boxplot prepared={
            lower whisker=0.106700211184998,
            lower quartile=0.115401890996878,
            median=0.13477684865461,
            upper quartile=0.145801552964923,
            upper whisker=0.175001045784812,
        },
    ] 
    coordinates{};
    
    \addplot+ [
        boxplot prepared={
            lower whisker=0.101526127298373,
            lower quartile=0.106666547279554,
            median=0.111040340107483,
            upper quartile=0.123342563063268,
            upper whisker=0.162581900373763,
        },
    ] 
    coordinates{};
    
    \addplot+ [
        boxplot prepared={
            lower whisker=0.08539983089383,
            lower quartile=0.100894764178075,
            median=0.106216826886396,
            upper quartile=0.115344911169959,
            upper whisker=0.198037614828185,
        },
    ] 
    coordinates{};
        
    \end{axis}
    \end{tikzpicture}
    
    
    \caption{Values of $\varepsilon$ for the group of women aged $19$--$89$ shown in increasing order of age, after post-processing according to cutting out trends with pattern $(3,3)$.
    Depiction of statistical quantities ($0$--$4$th quartile) as described in caption of Fig \ref{fig:womenFull}.
    }
    \label{fig:women_after_cut2}
\end{figure}

\begin{figure}[htbp]
    \centering
    \begin{tikzpicture}
    \begin{axis}[boxplot/draw direction=y,x axis line style={opacity=0},axis x line*=bottom,axis y line=left,enlarge y limits,ymajorgrids,xtick={1,2,3,4,5,6,7},xticklabels={20, 30, 40, 50, 60, 70, 80},]
   
    \addplot+ [
        boxplot prepared={
            lower whisker=0.122856960597144,
            lower quartile=0.125556810884963,
            median=0.143533178110005,
            upper quartile=0.170326216177691,
            upper whisker=0.208381244480661,
        },
    ] 
    coordinates{};

    \addplot+ [
        boxplot prepared={
            lower whisker=0.11521011186369,
            lower quartile=0.127624476954781,
            median=0.137163505432961,
            upper quartile=0.147643349647943,
            upper whisker=0.206767652335492,
        },
    ] 
    coordinates{};

    \addplot+ [
        boxplot prepared={
            lower whisker=0.111082200433754,
            lower quartile=0.129260139157375,
            median=0.140601442311665,
            upper quartile=0.158085592651659,
            upper whisker=0.1833816263911,
        },
    ] 
    coordinates{};
    
    \addplot+ [
        boxplot prepared={
            lower whisker=0.095920465284357,
            lower quartile=0.104539918792201,
            median=0.113159372300045,
            upper quartile=0.133071287624687,
            upper whisker=0.152983202949328,
        },
    ] 
    coordinates{};
    
    \addplot+ [
        boxplot prepared={
            lower whisker=0.083271422382505,
            lower quartile=0.128700800929701,
            median=0.133441137494413,
            upper quartile=0.145741573098049,
            upper whisker=0.167493546671243,
        },
    ] 
    coordinates{};
    
    \addplot+ [
        boxplot prepared={
            lower whisker=0.086519824405631,
            lower quartile=0.103313470020595,
            median=0.132215122363091,
            upper quartile=0.157734733638389,
            upper whisker=0.167300759977292,
        },
    ] 
    coordinates{};
    
    \addplot+ [
        boxplot prepared={
            lower whisker=0.107713608600343,
            lower quartile=0.11036763048189,
            median=0.111367510357994,
            upper quartile=0.11240390576241,
            upper whisker=0.121819178977675,
        },
    ] 
    coordinates{};
        
    \end{axis}
    \end{tikzpicture}
    
    \caption{Values of $\varepsilon$ for the group of men aged $21$--$88$ shown in increasing order of age, after post-processing according to cutting out trends with pattern $(3,3)$.
    Depiction of statistical quantities ($0$--$4$th quartile) as described in caption of Fig \ref{fig:womenFull}.}
    \label{fig:men_after_cut2}
\end{figure}
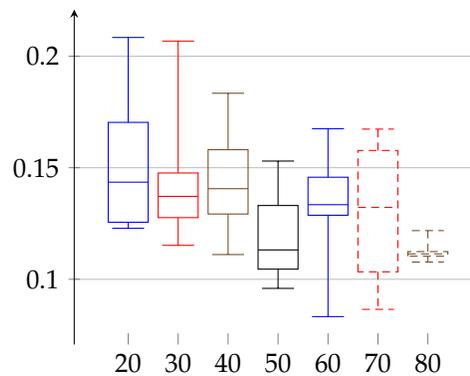

\subsection{Identifying \texorpdfstring{$\varepsilon$}{epsilon} for exemplary manually pre-processed data for medical purpose}\label{sec:med}

Although we concentrate on the cryptographical 
aspect of the heart-rate data, we extend our
findings to the ones with strictly medical meaning.
This is because the $\epsilon$-SV can, in principle, be treated as a parameter with potential novel meaning from the medical point of view. To exactly confirm its relevance, one should compare its value with the one obtained from persons with particular diseases. Here we show the values of $\tilde{\epsilon}$ for healthy persons.

\begin{figure}[htbp]
    \centering
    \begin{tikzpicture}
    \begin{axis}[boxplot/draw direction=y,x axis line style={opacity=0},axis x line*=bottom,axis y line=left,enlarge y limits,ymajorgrids,xtick={1,2,3,4,5,6,7,8},xticklabels={10,20, 30, 40, 50, 60, 70, 80},]
    \addplot+ [
        boxplot prepared={
            lower whisker   =0.210402967590354,
            lower quartile  =0.23214992306294,
            median          =0.253896878535526,
            upper quartile  =0.258779445761361,
            upper whisker   =0.263662012987197,
        },
        ] 
        coordinates{};
      
    \addplot+ [
        boxplot prepared={
            lower whisker   =0.189897291742424,
            lower quartile  =0.241657791931309,
            median          =0.255272408132298,
            upper quartile  =0.267225005017338,
            upper whisker   =0.283515472681021,
        },
        ] 
        coordinates{};

    \addplot+ [
        boxplot prepared={
            lower whisker   =0.214239445309983,
            lower quartile  =0.235181660164896,
            median          =0.238439319260143,
            upper quartile  =0.269039013532664,
            upper whisker   =0.285878663615177,
        },
        ] 
        coordinates{};

    \addplot+ [
        boxplot prepared={
            lower whisker   =0.209041757068668,
            lower quartile  =0.224137969605279,
            median          =0.265217371124695,
            upper quartile  =0.275554689782966,
            upper whisker   =0.292183588495742,
        },
        ] 
        coordinates{};
    
    \addplot+ [
        boxplot prepared={
            lower whisker   =0.19577230059331,
            lower quartile  =0.225895577815859,
            median          =0.250368851132734,
            upper quartile  =0.262225478467095,
            upper whisker   =0.307867973914234,
        },
        ] 
        coordinates{};
    
    \addplot+ [
        boxplot prepared={
            lower whisker   =0.210865176425069,
            lower quartile  =0.226505465040456,
            median          =0.262146777806612,
            upper quartile  =0.271868005355135,
            upper whisker   =0.299368075460046,
        },
        ] 
        coordinates{};
    
    \addplot+ [
        boxplot prepared={
            lower whisker   =0.184908917970767,
            lower quartile  =0.215339890393349,
            median          =0.261864601876293,
            upper quartile  =0.297500406629935,
            upper whisker   =0.371939609709359,
        },
        ] 
        coordinates{};
    
    \addplot+ [
        boxplot prepared={
            lower whisker   =0.175731971156014,
            lower quartile  =0.246373721656481,
            median          =0.279579880379113,
            upper quartile  =0.293333145789255,
            upper whisker   =0.33397514399303,
        },
        ] 
        coordinates{};
        
    \end{axis}
    \end{tikzpicture}
    
    
    \caption{Values of $\varepsilon$ for the group of women aged $19$--$89$ shown in increasing order of age, after manual pre-processing by a medical expert.
    Depiction of statistical quantities ($0$--$4$th quartile) as described in caption of Fig \ref{fig:womenFull}.
    }
    \label{fig:womencut500}
\end{figure}
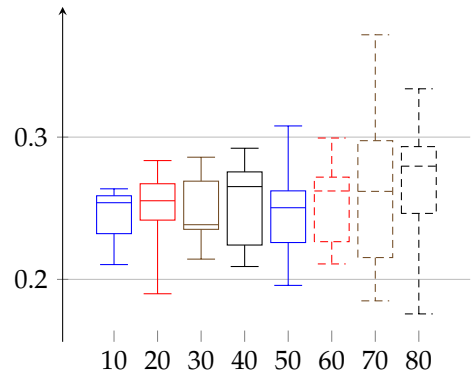

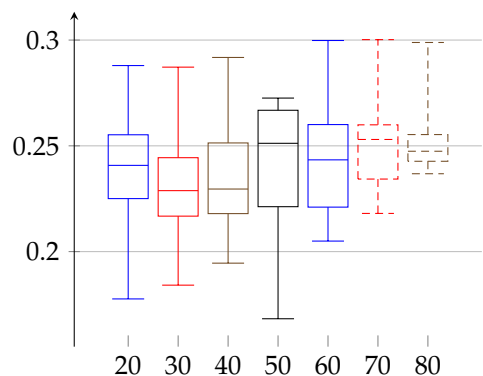
\begin{figure}[htbp]
    \centering
    \begin{tikzpicture}
    \begin{axis}[boxplot/draw direction=y,x axis line style={opacity=0},axis x line*=bottom,axis y line=left,enlarge y limits,ymajorgrids,xtick={1,2,3,4,5,6,7},xticklabels={20, 30, 40, 50, 60, 70, 80},]

    \addplot+ [
        boxplot prepared={
            lower whisker   =0.177632616627571,
            lower quartile  =0.225054154014819,
            median          =0.240765768137319,
            upper quartile  =0.255228838560659,
            upper whisker   =0.287943620194161,
        },
        ] 
        coordinates{};

    \addplot+ [
        boxplot prepared={
            lower whisker   =0.184141177021861,
            lower quartile  =0.216764467716763,
            median          =0.228817727051306,
            upper quartile  =0.244392298378604,
            upper whisker   =0.287198683735189,
        },
        ] 
        coordinates{};

    \addplot+ [
        boxplot prepared={
            lower whisker   =0.194540624900445,
            lower quartile  =0.217961380012618,
            median          =0.229577323800879,
            upper quartile  =0.251367566805995,
            upper whisker   =0.29177775217209,
        },
        ] 
        coordinates{};
    
    \addplot+ [
        boxplot prepared={
            lower whisker   =0.168257310561577,
            lower quartile  =0.221231842016988,
            median          =0.251196404126372,
            upper quartile  =0.266812314028728,
            upper whisker   =0.272576242783888,
        },
        ] 
        coordinates{};
    
    \addplot+ [
        boxplot prepared={
            lower whisker   =0.204974455007091,
            lower quartile  =0.221025950665102,
            median          =0.243371462715341,
            upper quartile  =0.260048878296471,
            upper whisker   =0.299792538526932,
        },
        ] 
        coordinates{};
    
    \addplot+ [
        boxplot prepared={
            lower whisker   =0.218049984968576,
            lower quartile  =0.234293694807568,
            median          =0.253010579850817,
            upper quartile  =0.259928098280382,
            upper whisker   =0.30020300268767,
        },
        ] 
        coordinates{};
    
    \addplot+ [
        boxplot prepared={
            lower whisker   =0.236809951646674,
            lower quartile  =0.242699896536332,
            median          =0.247440883791246,
            upper quartile  =0.255323780453711,
            upper whisker   =0.298854593617476,
        },
        ] 
        coordinates{};
        
    \end{axis}
    \end{tikzpicture}
    
    \caption{Values of $\varepsilon$ for the group of men aged $21$--$88$ shown in increasing order of age, after manual pre-processing by a medical expert.
    Depiction of statistical quantities ($0$--$4$th quartile) as described in caption of Fig \ref{fig:womenFull}.}
    \label{fig:mencut500}
\end{figure}

In this medical approach, we have studied manually pre-processed data of 190 persons. The first group consists of $88$ women aged $19$-$89$ years and the second consists of $102$ men aged $21$-$88$ years. We divided each group into subgroups of persons with age belonging to interval $[10*j,10*(j+1)]$ for $j\in\{1,...,8\}$ for women and $[10*k,10*(k+1)]$ with $k \in \{2,...,8\}$ for men. 
The age groups consisted of 10: 3, 20: 15, 30: 11, 40: 13, 50: 13, 60: 12, 70: 10, and 80: 11 elements in case of women and 20: 17, 30: 12, 40: 20, 50: 19, 60: 15, 70: 12, and 80: 7 in case of men.
On Figures \ref{fig:womencut500} and \ref{fig:mencut500} we describe the computed $\epsilon$ 
according to definition given in Eq.\ (\ref{eq:epsilon}) for each sex respectively. In the case of women the value of epsilon reads $\epsilon^{(w)} \in [0.17573, 0.33397]$ and in case of men the epsilon is in the range $\epsilon^{(m)} \in [0.16825, 0.30020]$, what resembles the results obtained from cryptographic signals, but there is no observed age dependency on the age group medians.

\subsection{Comparison of different experimental methods}\label{sec:comparison}

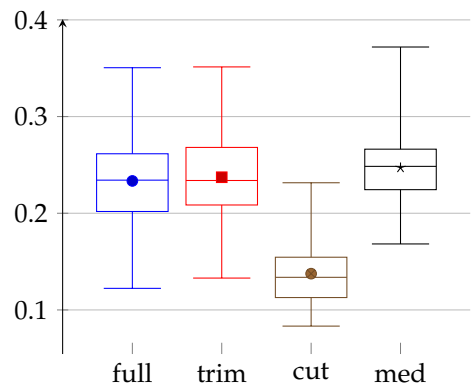
\begin{figure}[htbp]
    \centering
        \begin{tikzpicture}
    \begin{axis}[boxplot/draw direction=y,x axis line style={opacity=0},axis x line*=bottom,axis y line=left,enlarge y limits,ymajorgrids,xtick={1,2,3,4},xticklabels={full, trim, cut, med},]
    
    \addplot+ [
        boxplot prepared={
            lower whisker   =0.122324589075867,
            lower quartile  =0.201823507158334,
            median          =0.234236477976631,
            upper quartile  =0.261523334007569,
            upper whisker   =0.350524603419143,
        },
        ] table[row sep=\\,y index=0] {data\\ 0.233420725235202\\};
        
    \addplot+ [
        boxplot prepared={
            lower whisker   =0.132951832681267,
            lower quartile  =0.208550641968188,
            median          =0.233881928294179,
            upper quartile  =0.268086301407399,
            upper whisker   =0.351356813229199,
        },
        ] table[row sep=\\,y index=0] {data\\ 0.237243045360125\\};
        
    \addplot+ [
        boxplot prepared={
            lower whisker   =0.083271422382505,
            lower quartile  =0.112806370122659,
            median          =0.133729812346519,
            upper quartile  =0.154519852420001,
            upper whisker   =0.231539049293252,
        },
        ] table[row sep=\\,y index=0] {data\\ 0.137511622287952\\};
        
    \addplot+ [
        boxplot prepared={
            lower whisker   =0.168257310561577,
            lower quartile  =0.224397815901055,
            median          =0.248495038447646,
            upper quartile  =0.26627659987328,
            upper whisker   =0.371939609709359,
        },
        ] table[row sep=\\,y index=0] {data\\ 0.247030003564216\\};

    \end{axis}
    \end{tikzpicture}
    \caption{Collective values of $\varepsilon$ for all persons (regardless of age and sex) for four experiments presented in our work. For description see Section \ref{sec:comparison} 
    Depiction of statistical quantities ($0$--$4$th quartile) as described in caption of Fig \ref{fig:womenFull}. Additionally, bold dots in the figure represent average values.}
    \label{fig:comparison}
\end{figure}

To conclude the main part of our experiment, we present in Figure \ref{fig:comparison} comparison of different methods that we have used during our studies. In this figure, we analyze four different approaches: full - referring to the full data, trim - full data trimmed to the same length, cut - full data after ``cutting out trends'', and med - manually pre-processed data. The first three are discussed in Section \ref{sec:crypto}. The fourth one is described in Section \ref{sec:med}. 

In this comparison, we can make few interesting observations.
Firstly, trimming data to the same length have a marginal impact on the values of epsilons. It may be due to the fact that this trimming removes only a small portion of data. 
Secondly, epsilons for medical data do not differ much from the one obtained in the ``full'' approach. On one hand, in the medical case, the amount of data (for each person) is in fact a few times smaller, but on the other hand, the data are carefully chosen (both by taking into consideration only sleep time and manually removing incorrect values).
Finally, the ``cut'' approach, which uses ``cutting out trends'' gives significantly better results. It, therefore, shows that the carefully chosen pre-processing can impact the quality of obtained randomness, however at a price of a lower amount of data. Our ``cutting out trends'' pre-processing is successful due to the fact that the heartbeat has natural periodicity built in, which is obviously predictable, i.e., not random with respect to the adversary. 

\section{Discussion}\label{sec:discusion}

In this manuscript, we have considered the heart rate as a weak source of randomness in the context of quantum methods of its amplification. Since one of these methods is an amplification of the $\varepsilon$-Santha-Vazirani source, we have checked if the heart rate can be modeled as such a source. We have proposed the way to attribute $\varepsilon$ to \textit{finite} data according to which longer histories have a lower weight than shorter ones.
Since this proposition is heuristic,  other ways could be considered.
We have also proposed pre-processing of the signal, called cutting trends, which is natural in the context of heart rate data. It decreased the value of epsilon by half. It is interesting if other pre-processing or discretization of the signal can lead to lower values of $\varepsilon$.

In the experimental part of this manuscript, we have considered particular data and estimated its $\varepsilon$. Interestingly, its value does not strongly depend neither on the age nor sex of voluntaries and after cutting trends it is of the order of $\approx 0.13$. It seems then to be a parameter of a heart as a human's muscle.
One can also consider a heart-based Santha-Vazirani parameter $\varepsilon\neq 1/2$ as \textit{a precondition of life} of every heart possessing being, including humans. 

Fitting a quantum randomness amplification scheme to the above result would be the next step towards a device
which is not vulnerable to attack based on the correlation between source and device \cite{Max_Rotem, Wojewodka2017}.

From the medical point of view, the values of $\varepsilon$ obtained from the data with physiological meaning do not differ much
from the raw-data derived one. 
Moreover, the observed lack of dependence of $\epsilon$ on age suggests that there could be a universal pattern of a healthy heart rhythm. Therefore, it would be interesting in future to compare the value of $\varepsilon$ obtained from healthy persons which we have shown, with the one from persons
with particular diseases. This would give additional medical meaning to a cryptographic parameter leading to an interesting interplay between the two apparently far domains. 
Further, studying the pattern of the cut out trend (in our case, from the cryptographic point of view the most successful was (3,3)), can be of further interest also from the medical point of view.





\vspace{6pt} 



\authorcontributions{
    Conceptualization, K.H.; Methodology, M.S., K.H, and D.M.; Software, M.S, O.S., and K.H.; Formal Analysis, M.S., K.H., and D.M.; Investigation, M.S. and K.H.; Resources, D.M.; Data Curation, M.S. and O.S.; Writing – Original Draft Preparation, M.S. and K.H; Writing – Review \& Editing, M.S., K.H., and D.M.; Visualization, M.S.; Supervision, K.H. and D.M. All authors have read and agreed to the submitted version of the manuscript.
}

\funding{
    M.S., K.H., and O.S. acknowledge partial support by the National Science Centre grant Sonata Bis 5 no.\ 2015/18/E/ST2/00327. 
    M.S. acknowledges the National Science Centre, Poland, grant SHENG 1 no.\ 2018/30/Q/ST2/00625.
    K.H. acknowledges partial support by the Foundation for Polish Science (IRAP project, ICTQT, contract no.\ 2018/MAB/5, co-financed by EU via Smart Growth Operational Programme).
    }

\institutionalreview{
    The current calculations are based on the series presented in \cite{Makowiec.2015Frontiers} and \cite{Makowiec.2019Entropy}, i.e., on the signals collected for the research, which was approved by the Bioethical Committee of the Medical University of Gdańsk, as carried out in accordance with the Helsinki Declaration (Decision numbers: NKEBN/142/2009 and NKBBN/142-653/2019).
}

\informedconsent{
    All participants of the study gave their written consent to participate in those studies.
}

\dataavailability{
    The spreadsheet containing all results obtained from our program and further calculations is available in supplementary materials of this work (for details see Appendix \ref{sec:spreadsheet}). 
	Source heart rate data files can be sent upon request. Our program is closed source software.
} 

\acknowledgments{
    Authors acknowledge 
        Marta Żarczyńska-Buchowiecka 
        and 
        Joanna Wdowczyk 
        from the Medical University of Gdańsk 
        for sharing the data.
    The authors would like to thank 
    	Sreetama Das
    	for help with the implementation of the ``cutting out trends'' subroutine.
}

\conflictsofinterest{
    The authors declare no conflict of interest.
} 





\appendixtitles{yes} 

\appendixstart

\appendix

\section{Detailed numerical results}\label{sec:spreadsheet}

In this section, we will describe the form of detailed results of the numerical experiments. For the sake of the reader, we provide the results in a form of a spreadsheet attached to this paper as supplementary materials. The spreadsheet file named ``results.ods'' is divided into several tabs describing different approaches which we used.

In the first of the four tabs in the file, we present results for all persons separately in different settings. At the top, each column represented one person what is indicated by the corresponding name of the input data file. In these tabs, for the convenience of the reader, we used the following color-coding. In yellow, there are input file names, input file sizes, and numbers of data entries obtained from the input files after preprocessing (number of bits used in randomness test). Below that, in cyan, there are all epsilons estimated by our SVTest program, where each row corresponds to a different length of history used in the test. Then, in green, there are single values of epsilons for each person calculated according to Eq.\ (\ref{eq:epsilon}). At the bottom, in red, we presented various statistics for sets of persons grouped by sex and age intervals. The statistics are described in the main text and depicted in the appropriate figures. Additionally, texts in bold font are header describing different fields. Finally, all other values, with white background, are local variables used in calculations and can be omitted (unless the reader wants to fully understand details of the statistics calculations).

In Tab 1 we used full data files for each person using only necessary pre-processing without cutting out trends. Statistics for this approach are presented in Figures \ref{fig:womenFull} and \ref{fig:menFull}.

In Tab 2 we used the same data as above but we trimmed the number of entries to be the same for each person. This allows us to analyze data without getting into consideration potential epsilons bias that can be a result of different numbers of bits. Statistics for this approach are presented in Figures \ref{fig:womenNormalized} and \ref{fig:menNormalized}.

In Tab 3 we applied cutting out trends procedure (using parameters (3,3)) to the full data from Tab 1. Statistics for this approach are presented in Figures \ref{fig:women_after_cut2} and \ref{fig:men_after_cut2}.

In Tab 4 we present results for the medical approach that involves manual pre-processing. Appropriate statistics are presented in Figures \ref{fig:womencut500} and \ref{fig:mencut500}.

In Tab 5 we present results, in various approaches (cutting out trends with different parameters and original file trimmed to analogical lengths), obtained from the large input files consisting of data from all persons concatenated together. This approach allows us to simulate the behavior of the heart rate in a very long (possibly multi-day) measurement. Part of these results is summarized in Table \ref{tab:epsilons}.

In Tab 6 we show results for different approaches (similar to tab 5) for a single person.

Finally, In Tab 7 we show collective results of all persons (from all age groups and both sex) for different approaches. Then numbers 1 to 4 corresponds to Tabs 1 to 4 from above. Statistics for this approach are presented in Figure \ref{fig:comparison}.



\end{paracol}

\reftitle{References}



\externalbibliography{yes}
\bibliography{HeartSVSource}

\end{document}